# A Comparative Study of Single-Constraint Routing in Wireless Mesh Networks Using Different Dynamic Programming Algorithms

**Sabreen Mahmood Shukr**
(Master of Science)
Department of Computer Engineering
College of Engineering
University of Baghdad
Email: msh2008msha@yahoo.com

**Nuha Abdul Sahib Alwan**
Assistant Professor
Department of Computer Engineering
College of Engineering
University of Baghdad
Email: n.alwan@ieee.org

**Ibraheem Kassim Ibraheem**
Assistant Professor, Ph.D.
Department of Electrical Engineering
College of Engineering
University of Baghdad
Email: ibraheem151@gmail.com

**ABSTRACT:**

Finding the shortest route in wireless mesh networks is an important aspect. Many techniques are used to solve this problem like dynamic programming, evolutionary algorithms, weighted-sum techniques, and others. In this paper, we use dynamic programming techniques to find the shortest path in wireless mesh networks due to their generality, reduction of complexity and facilitation of numerical computation, simplicity in incorporating constraints, and their conformity to the stochastic nature of some problems. The routing problem is a multi-objective optimization problem with some constraints such as path capacity and end-to-end delay. Single-constraint routing problems and solutions using Dijkstra, Bellman-Ford, and Floyd-Warshall algorithms are proposed in this work with a discussion on the difference between them. These algorithms find the shortest route through finding the optimal rate between two nodes in the wireless networks but with bounded end-to-end delay. The Dijkstra-based algorithm is especially favorable in terms of processing time. We also present a comparison between our proposed single-constraint Dijkstra-based routing algorithm and the mesh routing algorithm (MRA) existing in the literature to clarify the merits of the former.

<div dir="rtl">

## مقارنة دراسية لتحديد المسار بمحدد واحد في الشبكات اللاسلكية المعشقة باستخدام مختلف خوارزميات البرمجة الديناميكية

صابرين محمود شكر
قسم هندسة الحاسبات
كلية الهندسة – جامعة بغداد

أ.م. نهى عبد الصاحب العلوان
قسم هندسة الحاسبات
كلية الهندسة – جامعة بغداد

أ.م.د. ابراهيم قاسم ابراهيم
قسم الهندسة الكهربائية
كلية الهندسة – جامعة بغداد

**الخلاصة:**

العثور على الطريق الاقصر في الشبكات اللاسلكية المعشقة هو امر هام. العديد من التقنيات قد استخدمت على حل هذه المشكلة مثل البرمجة الديناميكية، الخوارزميات التطورية، وتقنية المجموع الموزون للمحددات، وغيرها. في هذا البحث، استخدمنا تقنية البرمجة الديناميكية لإيجاد الطريق الاقصر الشبكات اللاسلكية المعشقة بسبب عموميته، والحد من التعقيد وتيسير الحساب العددي، بساطة في دمج المحددات، ومطابقته للطبيعة العشوائية لبعض المشاكل. مشكلة التوجيه هي مشكلة تحسين اهداف متعددة مع بعض المحددات مثل سعة المسار و الوقت من بداية المسار إلى نهايته. مشكلة التوجيه بمحدد واحد في خوارزميات Dijkstra، Bellman-Ford، و- Floyd

</div>

**Sabreen Mahmood Shukr**
**Nuha Abdul Sahib Alwan**
**Ibraheem Kassim Ibraheem**

**A Comparative Study of Single-Constraint Routing in Wireless Mesh Networks Using Different Dynamic Programming Algorithms**

Warshall والفرق فيما بينهم قد عرض في هذا العمل. هذه الخوارزميات تجد الطريق الاقصر من خلال ايجاد السعة المثلى بين عقدتين في الشبكات اللاسلكية ولكن مع تحديد الوقت الذي تحتاجه من بداية المسار الى نهايته. يتميز خوارزمي Dijkstra بقصر وقت المعالجة. وقد تناولنا أيضا المقارنة بين خوارزمياتنا وخوارزمية MRA المستحدثة سابقا وبيان فضل الأولى.

**الكلمات الرئيسية:** الشبكات اللاسلكية المعشقة، البرمجة الديناميكية، محدد واحد، خوارزمي Dijkstra، خوارزمي Bellman-Ford، خوارزمي Floyd-Warshall، خوارزمي MRA.

## 1. INTRODUCTION

Wireless mesh network (WMNs) are an attractive technology because they extend the network coverage, deliver community broadband Internet access services and increase the capacity of wireless access networks, **Cheng, H. et al., 2012, Capone A. et al., 2010, and Mountassir T. et al., 2012.** WMNs also reduce the need for costly wired network infrastructures, **Capone A. et al., 2010, and Marina M. K. et al., 2010.** Nodes in these networks establish and maintain connectivity amongst them automatically, **Raja N. K. et al., 2012.** The coverage of the network depends on the number of nodes, their location in relation to one another and the radio technology used because each node exchanges routing information only with its neighbors, **Kowalik K. et al., 2006.** To speed up the operation of finding the optimal path between two nodes in the WMNs, the dynamic programming technique is employed.

Dynamic programming (DP) is a powerful algorithmic paradigm which is used to solve large classes of optimization problems, **Lew A. et al., 2007.** The key attribute that a problem must have in order for dynamic programming to be applicable is optimal substructure, **Camen T. et al., 2001.** Optimal substructure means that the solution to a given optimization problem can be obtained by a combination of optimal solutions to its sub-problems. DP has many applications such as integer knapsack problem, optimal linear search problem, optimal binary search tree problem, integer linear programming, finding the shortest path problem and others, **Lew A. et al., 2007**. In this work, DP is used to find shortest paths problem in wireless networks in the context of quality of service (QoS) routing. The importance of focusing on DP techniques for routing stems from the fact that DP has been proven to be effective for many unconstrained and multi-objective optimization problems. For some problems, it is even the most efficient approach known for the solution. Wireless networks can be unidirectional or bidirectional mesh wireless networks. Each type of wireless network can make use of routing algorithms under the framework of dynamic programming techniques to determine the shortest paths between the nodes.

In unidirectional wireless networks, networks can be represented as a multi-stage directed acyclic graph (DAG). There are two fundamental processes to solve the shortest path problems in multi-stage directed graphs, **Moon T. K. and Stirling W. C., 2000.** The first one is known as forward dynamic programming algorithm and the other is known as backward dynamic programming algorithm. For wireless mesh networks, there are other applicable unconstrained DP algorithms such as the Dijkstra, Bellman-Ford and Floyd-Warshall techniques, **Stallings W., 2011, Glisic S. and Lorenzo B., 2009.** The processing time and the amount of information that must be collected from other nodes, **Stallings W., 2011** are the important aspects in which these algorithms differ from each other. Unconstrained shortest path problems and single-constraint optimization for path selection have been proved to be NP-complete, because they are all solved in polynomial time, **Dasgupta S. et al, 2007, and Allard G. et al., 2004.** A problem is NP-complete if all decisions for the problem can be verified in polynomial time, **Dasgupta S. et al., 2007.**

The routing problem is a multi-objective optimization problem with metrics such as path rate or capacity, end-to-end delay, hop count, and probability of errors. These QoS requirements need to be bounded or optimized.

In general, the QoS metrics need to be optimized by making meaningful trade-offs due to their inter-conflicting nature.

In this paper, we address the single-constraint routing problem in Dijkstra-based, Bellman Ford-based and Floyd Warshall-based DP framework. These algorithms find the shortest path from the source node to the destination node in the wireless network through maximizing path rate and limiting end-to-end delay. Also, we show that our Dijkstra-based algorithm is preferable over the mesh routing algorithm (MRA) in finding the shortest path in wireless mesh networks. MRA, **Crichigno J. et al., 2008, and Helonde, J. B. et al., 2011,** is also a dynamic programming algorithm to compute high-capacity end-to-end delay bounded paths. There are numerous other optimization methods each with advantages and disadvantages. The most prominent is dynamic programming due to its generality, reduction of complexity and facilitation of numerical computation, simplicity in incorporating constraints, and its conformity to the stochastic nature of some problems, **Lew A. and Mauch H., 2007, Doerr B. et al, 2009, and Ilaboya R. et al., 2011.** Evolutionary algorithms such as the genetic algorithm (GA) are most appropriate for complex non-linear models where the location of a global minimum is a difficult task. Due to global search, GAs are computationally expensive and need more time to be computed as compared with DP, **Doerr B. et al., 2009.**

The remainder of the paper is organized as follows: Section 2 presents the DP algorithms (MRA, Dijkstra, Bellman-Ford, and Floyd-Warshall) that deal with single constraints to find the shortest path in wireless mesh networks. Section 3 summarizes the simulation results and discussion. Finally, Section 5 concludes the paper.

## 2. SINGLE-CONSTRAINT DP ALGORITHMS

In this section, four single-constraint DP algorithms (MRA, Dijkstra, Bellamn-Ford, and Floyd-Warshall) are explained. All these algorithms follow the same method to find the shortest path in wireless networks. The method optimizes the path capacity and bounds the end-to-end delay. This method is designed to dispense with combined metrics. Combined metrics (weighted sum techniques) are optimized when the end-to-end delay metric is minimized and the path capacity is maximized simultaneously. The weighted sum technique does not guarantee any optimal trade-off solution between the metrics. Hence, with the weighted sum technique, solutions that best satisfy QoS requirements are not guaranteed, **Crichigno J. et al, 2008.**

In what follows, we explain the difference between the MRA algorithm and our single-constraint Dijkstra-based algorithm, **Shukr S. et al, 2012** and also explain the single-constraint Bellman-Ford and Floyd-Warshall algorithms.

### 2.1 The MRA Algorithm

The mesh routing algorithm (MRA), **Crichigno J. et al, 2008 and Helonde, J. B. et al., 2011** is a dynamic programming approach to compute high-capacity end-to-end delay bounded paths.

**Fig. 1** is used to show how the algorithm operates, where rate r(*l*) is shown over each link in Mbps, and the delay of each link t(*l*) equal to 2ms.

Now, we want to find the path from u to y with maximum capacity R and with delay exactly equal to 6ms ($\tau$= 3 hops time). This path is denoted by $P_6^*(u,y)$. Only through w or x, we reach y. The path rate from u to y is:

$$R(P_6^*(u,y)) = \max \begin{Bmatrix} \min\{R(P_{6-2}^*(u,w)),6\} \\ \min\{R(P_{6-2}^*(u,x)),3\} \end{Bmatrix} \quad (1)$$

The two paths ($P_{6-2}^*(u,w)$ and $P_{6-2}^*(u,x)$) can be similarly found. **Fig.2** explains how to solve the shortest path problem in **Fig.1** using the MRA algorithm. After we search all possible paths and discard some of them like *P3* and *P6* because their end-to-end delays greater than 6ms, we choose *P5* as the maximum path rate (5Mbps) with end-to-end delay equal to 6ms among all remaining paths.

The following pseudo-code shows how the MRA algorithm operates to find the maximum capacity path with end-to-end delay bounded by $\tau$.

**The MRA algorithm**

1. **INPUT:** *G(V,E)*, source node vs, destination node vd, delay bound $\tau$.
2. **OUTPUT:** $P_{vd}^{vs}$

3. /*Initialization*/
4. **IF** pair u, w ϵ V and $d < \tau$ **THEN**
5.     $P_d^*(u,w)$ =NIL
6.     **IF** ∃ $l$ = (u,w) ϵ E | $t(l)=d$ **THEN**
7.         $P_d^*(u,w) = l$
8.     **END IF**
9. **END IF**
10. /*main loop*/
11. **IF** $d < \tau$ **THEN**
12.   **IF** pair u, w ϵ V **THEN**
13.     **FOR ALL** $l$=(u,v) ϵ E | $d > t(l)$ **DO**
14.       **IF** $P_{d-t(l)}^*(v,w) \neq$ NIL **THEN**
15.         **IF** [ $P_d^*(u,w) =$ NIL] or
    [ $R(P_d^*(u,w)) <$
    $R(l \oplus P_{d-t(l)}^*(v,w))$ ] **THEN**
16.           $P_d^*(u,w) = l \oplus P_{d-t(l)}^*(v,w)$
17.         **END IF**
18.       **END IF**
19.     **END FOR**
20.   **END IF**
21. **ENF IF**
22. **RETERN** $P_{vd}^{vs}$.

## 2.2 Dijkstra-Based Algorithm

We propose a Dijkstra DP technique that computes high-capacity paths while simultaneously bounding the end-to-end delay to an upper limit. **Fig.3** explains how the algorithm works.

Beginning with the source node (*vs*), the algorithm finds node (*u*) whose *R(P(vs,u))* is the maximum capacity among all nodes. After that the algorithm finds the links ($l_{uv}$) that connect *u* and *v* for all *v* provided that the delay from *vs* to *v* does not exceed $\tau$. So the path from *vs* to *v* is either $P(vs,v)$ or $P(vs,u) \oplus l_{uv}$ depending on which has the maximum capacity, and at the same time, does not violate the delay bound. The path from *vs* to *v* throught *u* has a rate equal to $min\{R(P(vs,u)), r(l_{uv})\}$. The paths $P(vs,u)$ and $P(vs,v)$ are not necessarily disjoint. If $R(P(vs,u) \oplus l_{uv}) > R(P(vs,v))$ then $P(vs,v)$ is $P(vs,u) \oplus l_{uv}$.

By the same way we can expand the path to reach the destination node (*vd*) and find $P_\tau^*(vs,vd)$ and $R(P_\tau^*(vs,vd))$ denoting the shortest path (greatest capacity) and its rate respectively.

In short, our proposed algorithm adopts unconstrained (capacity or rate) Dijkstra optimization but prunes off the paths that violate the delay bound resulting in a single-constraint optimization effect. This is achieved by comparing rates to decide between paths, and then comparing delays. The latter comparison may change the decisions resulting from the former.

To show the difference between this algorithm and MRA algorithm, **Fig.4** shows the spanning tree of the same network in **Fig.1** using Dijkstra-based algorithm. It is clear that the optimum path *P1* is rapidly discovered. From this example, it is clear that our proposed algorithm is better than MRA because MRA employs flooding in its search, but with optimization.

Moreover, our algorithm is different from the MRA in that it is Dijkstra-based and, therefore, retains all the corresponding advantages such as fast shortest-path determination, and having an order of $N^2$ ( N is the number of the nodes) rendering it efficient to use with relatively large networks. The algorithm in the following pseudo-code is used to return a maximum-capacity path such that the end-to-end delay is bounded by $\tau$, and its rate.

**The single-constraint Dijkstra-based algorithm**

1. **INPUT**: no. of nodes n, source node vs, destination node vd, delay bound $\tau$, t(l) for all l, r(l) for all l.
2. **OUTPUT**: $P_\tau^*(vs,,vd)$, $R(P_\tau^*(vs,,vd))$.
3. /*Initialization*/
4. **FOR** ALL nodes
5.     Visited nodes = NIL
6.     R(P(vs,node))=0
7.     D(P(vs,node))=∞
8.     Parent(node)= NIL
9. **END**
10. R(P(vs,vs))=∞
11. D(P(vs,vs))=0
12. **FOR** i=1: (n-1)
13.     **FOR** ALL visited nodes
14.         Rate(node) = R(P(vs,node))

| | |
|---|---|
| 15. | **END** |
| 16. | $r_{max}$ = max[rate(node)] |
| 17. | u=node corresponding to $r_{max}$. |
| 18. | /* u is the visited node */ |
| 19. | **FOR** ALL nodes v |
| 20. |     **IF** (t($l_{uv}$) + D(P(vs,u))) < τ **THEN** |
| 21. |         **IF** r($l_{uv}$) < R(P(vs,u)) **THEN** |
| 22. |             cap=r($l_{uv}$) |
| 23. |         **ELSE** cap= R(P(vs,u)) |
| **24.** |         **END IF** |
| 25. |         **IF** cap> R(P(vs,v)) OR D(P(vs,v)) > τ **THEN** |
| 26. |             R(P(vs,v)) = cap |
| 27. |             parent(v) = u |
| 28. |             /* P(vs,v) = P(vs,u) $\oplus l_{uv}$ */ |
| 29. |             D(P(vs,v)) = D(P(vs,u)) + t($l_{uv}$) |
| | **ENDIF** |
| 30. |     **ELSE IF** D(P(vs,v))> τ **THEN** |
| 31. |         R(P(vs,v)) = 0 |
| 32. |         D(P(vs,v)) = ∞ |
| **33.** |     **END IF** |
| **34.** | **END FOR** |
| **35.** | **END FOR** |
| 36. |     **IF** parent(vd) ≠ NIL **THEN** |
| 37. |         P(vs,v) = [vd] |
| 38. |         t = vd |
| 39. |         **WHILE** t ≠ vs |
| 40. |             p = parent(t) |
| 41. |             P(vs,vd) = [p P(vs,vd)] |
| 42. |             t = p |
| **43.** |         **END** |
| **44.** |     **END IF** |
| 45. | $P_\tau^*(vs,,vd)$ = P(vs,vd) |
| 46. | **RETURN** $P_\tau^*(vs,,vd)$, R($P_\tau^*(vs,,vd)$) |

### 2.3 Bellman-Ford Algorithm

In the following pseudo-code, Bellman-ford algorithm finds the optimal path from source node (s) to all other nodes in the network. It uses output information from rates matrix, introduced earlier in the previous sub-section. First, we have only the s node in the path. For all links in the network, and from s node, we will find all rates and the corresponding nodes and compare between them to find the maximum rate ($r1_{max}$) and its corresponding node (m). Then, in the same manner but from node m, we will find $r2_{max}$ and its corresponding node. Then $r1_{max}$ will compare with $r2_{max}$ and the smallest be the maximum rate from the s node to the third node in the path. We do all these steps with delay bounded by $\tau$. All above processes will be repeated until all nodes in the network are examined.

**The single-constraint Bellman-Ford based algorithm**

| | |
|---|---|
| 1. | **INPUT**: no. of nodes n, no. of links L, source node vs, destination node vd, delay bound τ, t(l) for all l, r(l) for all l, links between all nodes. |
| 2. | **OUTPUT**: $P_\tau^*(vs,,vd)$, R($P_\tau^*(vs,,vd)$). |
| 3. | /*Initialization*/ |
| 4. | **FOR** ALL nodes |
| 5. |     Visited nodes = NIL |
| 6. |     R(P(vs,node))=0 |
| 7. |     D(P(vs,node))=∞ |
| 8. |     Parent(node)= NIL |
| 9. | **END** |
| 10. | R(P(vs,vs))=∞ |
| 11. | D(P(vs,vs))=0 |
| 12. | **FOR** i=1: (n-1) |
| 13. |     **FOR** ALL visited nodes |
| 14. |         Rate(node) = R(P(vs,node)) |
| 15. |     **END FOR** |
| 16. |     $r_{max}$ = max[rate(node)] |
| 17. |     u=node corresponding to $r_{max}$. |
| 18. |     /* u is the visited node */ |
| 19. |     **FOR** ALL links j |
| 20. |         /*find the node v which is connected to node u through link j */ |
| 21. |         **FOR** ALL nodes v |
| 22. |             **IF** (t(luv) + D(P(vs,u))) < τ **THEN** |
| 23. |                 **IF** r(luv) < R(P(vs,u)) **THEN** |
| 24. |                     cap=r(luv) |
| 25. |                 **ELSE** cap= R(P(vs,u)) |
| 26. |                 **END IF** |
| 27. |                 **IF** cap> R(P(vs,v)) OR D(P(vs,v)) > τ **THEN** |
| 28. |                     R(P(vs,v)) = cap |
| 29. |                     parent(v) = u |

```
30.            /* P(vs,v) = P(vs,u) ⊕
                  luv */
31.            D(P(vs,v)) = D(P(vs,u))
                  + t(luv)
32.            ELSE IF D(P(vs,v))> τ
                  THEN
33.                R(P(vs,v)) = 0
34.                D(P(vs,v)) = ∞
35.            END IF
36.        END IF
37.    END FOR
38. END FOR
39. IF parent(vd) ≠ NIL THEN
40.    P(vs,v) = [vd]
41.    t = vd
42.    WHILE t ≠ vs
43.        p = parent(t)
44.        P(vs,vd) = [p P(vs,vd)]
45.        t = p
46.    END
47. END IF
48. END FOR
49. $P_\tau^*$(vs,,vd) = P(vs,vd)
50. RETURN $P_\tau^*$(vs,,vd) , R($P_\tau^*$ (vs,,vd)).
```

## 2.4 Floyd-Warshall Algorithm

Floyd-Warshall algorithm in the following pseudo-code determines the shortest path between all pairs of nodes in the network. Floyd-Warshall algorithm uses a set of nodes as intermediate nodes to find the routes between the nodes in the network. Suppose we wish to find the route from *vs* node to *vd* node in the network. First, there is only node 1 as intermediate node between node *vs* and node *vd*. $r_{s1}$ represent the rate between node *vs* and node 1 and $r_{1d}$ represent the rate between node 1 and node *vd*. The smallest between node $r_{s1}$ and $r_{1d}$ will be the maximum rate ($r1_{max}$) between node *vs* and node *vd* under there is only node 1 as intermediate node. Then, the above process will be repeated but with node 2 as intermediate node and with $r2_{max}$ as the maximum rate between node *vs* and node *vd* under node 2 as intermediate node. The rate $r1_{max}$ will be compared with $r2_{max}$ and the largest will be the maximum rate between node *vs* and node *vd* and so on. We do all these steps with delay bounded by $\tau$. All these processes will be repeated until the all optimum routes between all pairs in the network are found.

**The single-constraint Floyd-Warshall based algorithm**

```
1.  INPUT: no. of nodes n, no. of links L,
    source node vs, destination node vd,
    delay bound τ, t(l) for all l, r(l) for all l,
    links between all nodes.
2.  OUTPUT: $P_\tau^*$(vs,,vd) , R($P_\tau^*$ (vs,,vd)).
3.  /*Initialization*/
4.  FOR ALL nodes
5.      Visited nodes = NIL
6.      Parent(node)= NIL
7.  END
8.  FOR k=1: n
9.      FOR i=1: n
10.         FOR j=1: n
11.             IF node (i) = node (j) THEN
12.                 /* rate and delay
13.                 between node i and
                    j stay the same*/
14.             ELSE
15.                 IF R(i,k) ==0 THEN
16.                     /*there is no
                        change*/
17.                 END IF
18.                 IF R(k,j) ==0 THEN
19.                     /*there is no
                        change*/
20.                 END IF
21.                 IF D(i,k) + D(k,j) < τ
                    THEN
22.                     Cap= min(R(i,k) ,
                        R(k,j))
23.                     IF (cap > R(i,j)) OR
                        (D(i,j) > τ) THEN
24.                         IF parent(i,k) ==
                            NIL
25.                             parent(i,j) =k
26.                         ELSE
27.                             parent(i,j) =
                                parent(i,k)
28.                         END IF
29.                         R(i,j) = cap
30.                         /* P(i,j) = P(i,k) ⊕
                            P(k,j) */
31.                         D(i,j) = D(i,k) +
                            D(k,j)
32.                     END IF
```

```
33.             ELSE
34.                IF D(i,j) > τ THEN
35.                    R(i,j) = 0
36.                    D(i,j) = ∞
37.                    Parent(I,j) = NIL
38.                END IF
39.             END IF
40.          END IF
41.       END FOR
42.    END FOR
43. END FOR
44. IF (R(vs,vd) !=0) AND (D(vs,vd) !=∞)
    THEN
45.    P(vs,vd) = [vd]
46.    t = vd
47.    WHILE P(t,vs) != 0
48.           p = parent(t,vs)
49.           P(vs,vd) = [p P(vs,vd)]
50.           t = p
51.    END
52.    P(vs,vd) = [vs P(vs,vd)]
53. END IF
54. $P_\tau^*$(vs,,vd) = P(vs,vd)
55. RETURN $P_\tau^*$(vs,,vd) , R($P_\tau^*$(vs,,vd)).
```

## 3. RESULTS AND DISCUSSION

The simulation program to implement the comparison between the single-constraint DP algorithms (MRA, Dijkstra, Bellman-Ford, and Floyd-Warshall) was coded in Matlab 8.0. The resultant shortest route depends on link rate and end-to-end delay. The simulation model parameters are chosen as follows: Number of nodes in the simulated network= 50. Topology area: Nodes are distributed randomly on 1000*1000 $m^2$. This network topology ensures that the node coverage area is 200 *m.* Thus, some nodes may be in the coverage area of others.

**Figs. 5**, **6**, and **7** show the topology of the network and the shortest route in terms of maximum capacity in Dijkstra, Bellman-Ford, and Floyd-Warshall with a single constraint from node 1 to node 15, node 23 to node 24, and node 49 to node 50 respectively. All the routes in the three figures are bounded by a 50 *ms* delay. The route in **Fig.5** has a capacity of 7.8151 *Mbps* while the route in **Fig.6** has a capacity of 8.1306 *Mbps* and the route in **Fig. 7** has a capacity of 8.1634 *Mbps*.

As shown from **Figs. 5**, **6**, and **7**, there is no difference among the results of the three algorithms (Dijkstra, Bellman-Ford, and Floyd-Warshall) because all of them find the shortest route between two nodes in the network. However, these approaches differ in the amount of information that must be collected from other nodes in the network. In Bellman-Ford algorithm, the node must collect only the information from its neighbors and knowledge of its link costs, to update its costs and paths. While in the Dijkstra algorithm, the node must know the link costs of all links in the network. The information must be exchanged with all other nodes. Thus, Bellman-Ford is better than Dijkstra from this point of view.

As for the processing time of the algorithms, the processing time in the Dijkstra algorithm is $O(N^2)$, where N is the number of nodes in the network, whereas in Bellman-Ford, the processing time is O(NL), where L is the number of links in the network. The processing time in the Floyd-Warshall is $O(N^3)$. The Dijkstra algorithm is better than other two algorithms in the processing time feature.

## 4. CONCLUSION

Wireless mesh networks are likely to be the essence of future communication. Finding shortest paths in WMNs by optimizing some QoS metrics is very challenging. Because it is difficult to provide optimization for all the metrics used to solve the routing problem simultaneously, we use a technique that optimizes one QoS measure and bounds or constrains the other. This paper presents single-constraint shortest path problem in Dijkstra, Bellman-Ford, and Floyd-Warshall dynamic programming algorithms that optimize the path capacity and bound the end-to-end delay. Despite the fact that the simulation results of the three algorithms are the same but they are different in the processing time of the algorithms and the amount of information that must be collected from other nodes in the wireless network. From the perspective of processing time, the Dijkstra algorithm is the best because the processing time is $O(N^2)$, while in Bellman-Ford algorithm is O(NL) and in Floyd-Warshall is $O(N^3)$. Bellman-Ford is better than the other algorithms in the amount of information must be collected from the nodes because the node only collects the information from its neighbors. Another comparison was implemented between the MRA and single-constraint Dijkstra-based algorithms to show the difference between them. The result of this comparison explains that our proposed algorithm is better than the MRA algorithm in the search of the shortest path due to its comparatively reduced computational complexity, whereas MRA is more time-consuming due to its flooding-like search method.

## 6. SYMBOLS AND ACRONYMS

| Symbol | Meaning |
|---|---|
| $\sum$ | Summation Operator |
| $\oplus$ | Concatenation |
| $\exists$ | There Exist |
| \| | OR |
| $\infty$ | Infinity |
| $\tau$ | End-to-End Delay Bound |
| d, $vd$ | Destination Nodes in the Wireless Network |
| d | Delay of the Path So Far |
| D(P) | End-to-End Delay of Path P |
| E | Number of Edges in the Graph |
| G(V,E) | Graph with V Vertices and E Edges |
| l | Link in the Path P |
| $l_{uv}$ | Link Connecting Node u and Node v |
| L | Number of Links in the Wireless Network |
| m, u, v, w, x, y | Nodes in the Wireless Network |
| N | Number of Nodes in the Wireless Network |
| NIL | Nothing |
| P | Path |
| $P_{vd}^{vs}$ | Path Between *vs* and *vd* |
| $P_\tau^*(vs,,vd)$ | Maximum-Capacity Path Between *vs* and *vd* with End-to-End Delay Less Than $\tau$ |
| $r_{max}$ | Maximum Rate |
| r(*l*) | Rate of Link *l* |
| R(P) | Rate of Path P |
| s, *vs* | Source Node in the wireless Network |
| t(*l*) | Expected Delay of Link *l* |
| DAG | Directed Acyclic Graph |
| DP | Dynamic Programming |
| GA | Genetic Algorithm |
| MRA | Mesh Routing Algorithm |
| NP | Nondeterministic Polynomial Time |
| QoS | Quality of Service |
| WMN | Wireless Mesh Network |

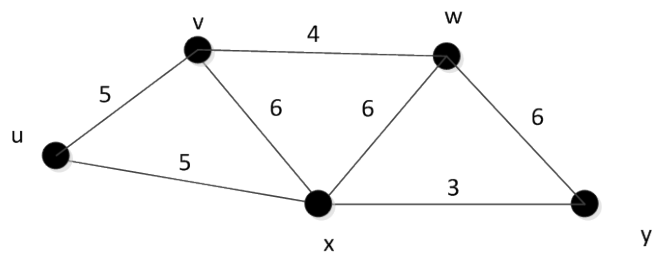

**Figure 1.** A network with rate r(*l*) over each link.

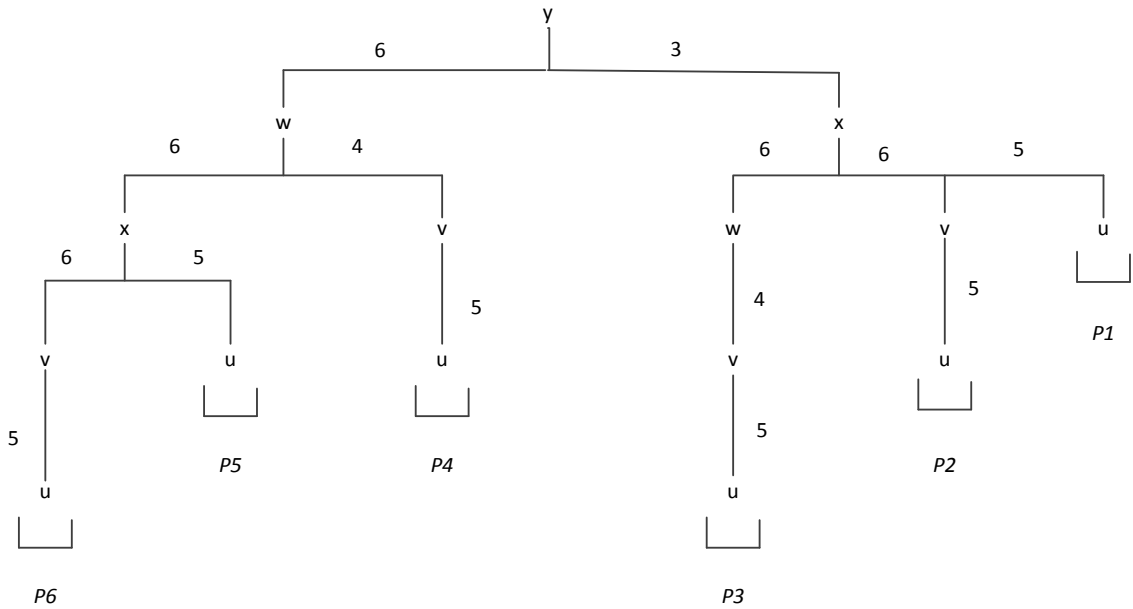

**Figure 2.** Spanning tree layout explains how to solve the shortest path problem of in Fig. 1 using MRA algorithm.

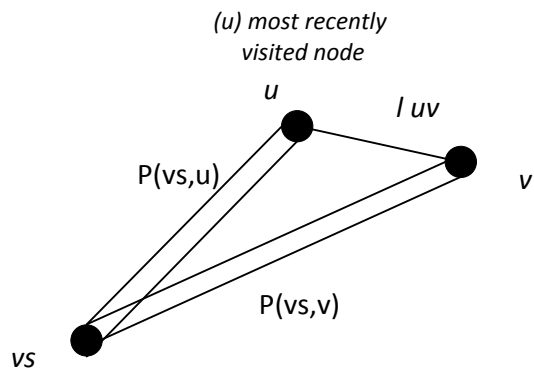

**Figure. 3:** The path from *vs* to *v*.

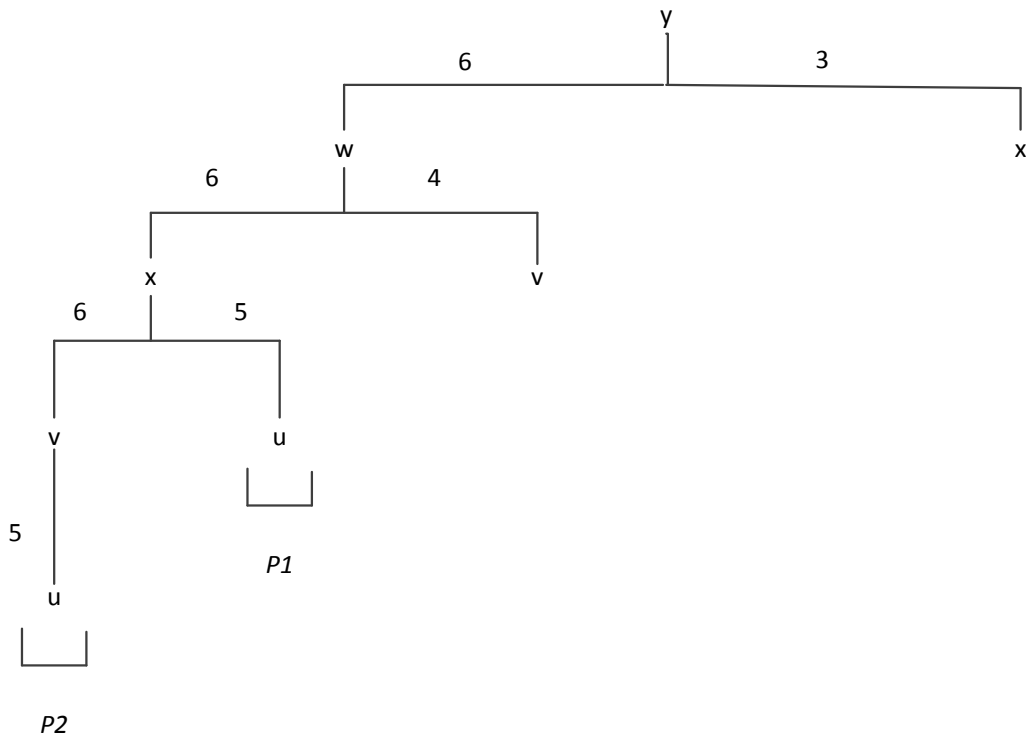

**Figure 4.** Spanning tree layout explains how to solve the shortest path problem of in Fig. 1 using Dijkstra-based algorithm.

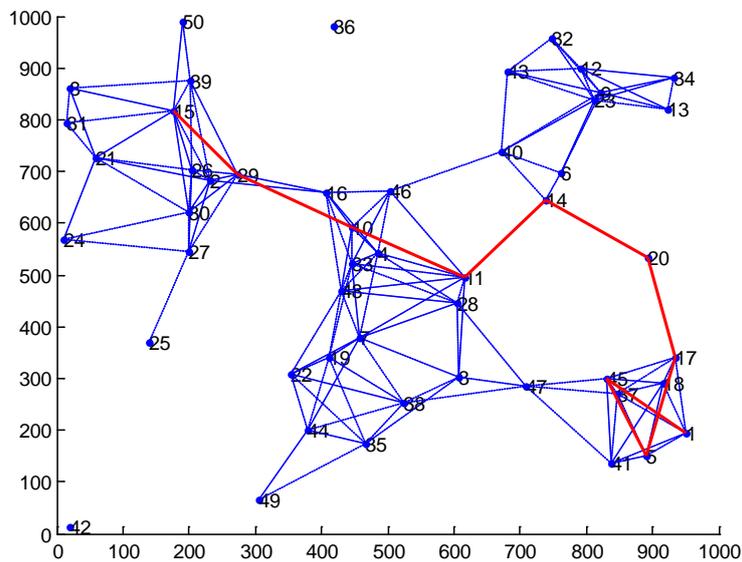

**Figure 5.** Network topology with source node (1) and destination node (15) showing shortest route under single- constraint DP algorithms with maximum capacity and a 50ms end-to-end delay bound.

**Sabreen Mahmood Shukr**
**Nuha Abdul Sahib Alwan**
**Ibraheem Kassim Ibraheem**

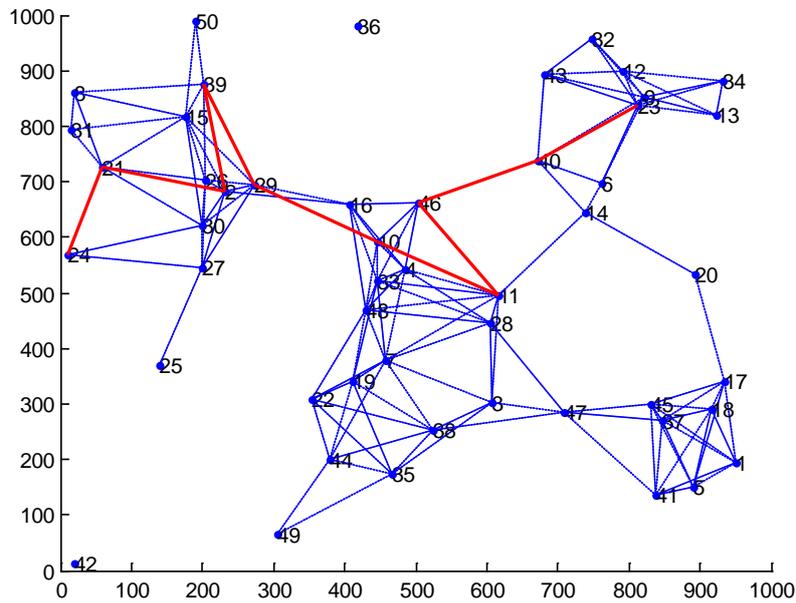

**Figure 6.** Network topology with source node (23) and destination node (24) showing shortest route under single- constraint DP algorithms with maximum capacity and a 50ms end-to-end delay bound.

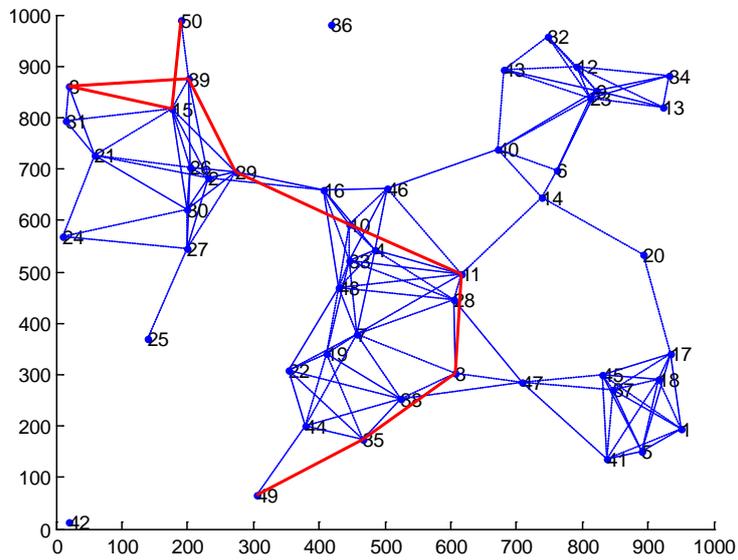

**Figure 7.** Network topology with source node (49) and destination node (50) showing shortest route under single- constraint DP algorithms with maximum capacity and a 50ms end-to-end delay bound.